\documentclass[5p,twocolumn]{elsarticle}

\usepackage{graphicx}
\usepackage{dcolumn}
\usepackage{pifont}
\usepackage{bm}
\usepackage{multirow}
\usepackage{amsmath}
\usepackage{float}
\usepackage{txfonts}
\usepackage[version=3]{mhchem}

\begin{document}
\title{First-principles study on the electronic and optical properties of inorganic perovskite \ce{Rb_{$1-x$}Cs_{$x$}PbI3} for solar cell applications}

\author[kimuniv-m,kimuniv-c]{Un-Gi Jong}
\author[kimuniv-m]{Chol-Jun Yu\corref{cor}}
\ead{ryongnam14@yahoo.com}
\author[kimuniv-m]{Yun-Sim Kim}
\author[kimuniv-m]{Yun-Hyok Kye}
\author[kimuniv-c]{Chol-Ho Kim}

\cortext[cor]{Corresponding author}

\address[kimuniv-m]{Computational Materials Design (CMD), Faculty of Materials Science, Kim Il Sung University, \\ Ryongnam-Dong, Taesong District, Pyongyang, Democratic People's Republic of Korea}
\address[kimuniv-c]{Natural Science Centre, Kim Il Sung University, Ryongnam-Dong, Taesong District, Pyongyang, Democratic People's Republic of Korea}

\begin{abstract}
Recently, replacing or mixing organic molecules in the hybrid halide perovskites with the inorganic Cs or Rb cations has been reported to increase the material stability with the comparable solar cell performance. In this work, we systematically investigate the electronic and optical properties of all-inorganic alkali iodide perovskites \ce{Rb_{$1-x$}Cs_{$x$}PbI3} using the first-principles virtual crystal approximation calculations. Our calculations show that as increasing the Cs content $x$, lattice constants, band gaps, exciton binding energies, and effective masses of charge carriers decrease following the quadratic (linear for effective masses) functions, while static dielectric constants increase following the quadratic function, indicating an enhancement of solar cell performance upon the Rb addition to \ce{CsPbI3}. When including the many-body interaction within the GW approximation and incorporating the spin-orbit coupling (SOC), we obtain more reliable band gap compared with  experiment for \ce{CsPbI3}, highlighting the importance of using GW+SOC approach for the all-inorganic as well as organic-inorganic hybrid halide perovskite materials.
\end{abstract}


\maketitle

\section{Introduction}
Recently, perovskite solar cells (PSCs) have emerged as the most promising photovoltaic (PV) solar cells, due to a low cost of raw materials, ease of manufacturing process~\cite{Liu,Burschka}, and moreover, rapidly evolving power conversion efficiency (PCE) over 22\%~\cite{WSYang}. In general, PSCs include the organic-inorganic halide perovskite as the major component with a chemical formula \ce{ABX3}, where A is an organic monovalent cation like methylammonium (MA = \ce{CH3NH3}) or formamidinium (FA = \ce{HC(NH2)2}), B is an inorganic divalent cation (Pb or Sn), and X is a halogen anion (I, Br, Cl). As a typical example, \ce{MAPbI3} possesses lots of good properties for solar absorber~\cite{Ganose}, such as appropriate band gaps for light absorption~\cite{Baikie}, low exciton binding energy~\cite{Innocenzo}, high charge carrier mobilities~\cite{Shao}, and tolerable defect properties~\cite{Walsh}. However, its poor material stability, especially under heat and humidity conditions, imposes restrictions on the long-term stability of PSCs, which is the most challenging obstacle to the large-scale commercial application of PSCs~\cite{Li,Berhe}. In the organic-inorganic hybrid perovskites, the organic molecules such as MA and FA are particularly sensitive to moisture and/or oxygen due to their high hygroscopicity, and thus it is not easy to prevent their decomposition leading to the degradation of PSCs upon exposure to moisture.

To resolve this issue, replacing the organic cations MA or FA with inorganic monovalent alkali cations such as caesium and rubidium has been suggested as a favourable way due to their less sensitivity to moisture~\cite{XLi,Walsh15}. Although the \ce{Cs+} (1.67 \AA) and \ce{Rb+} (1.52 \AA) cations have smaller ionic radius than \ce{MA+} (1.80 \AA) or \ce{FA+} (1.90 \AA), their Goldschmidt tolerance factors for \ce{APbI3} perovskites, defined as $\alpha=(r_{\ce{A}}+r_{\ce{X}})/\sqrt{2}(r_{\ce{B}}+r_{\ce{X}})$ with the ionic radii $r$, are 0.81 and 0.78~\cite{Saliba2}, within the proper range 0.7 $< \alpha <$ 1 for stable halide perovskites~\cite{Goldschmidt,Shannon,Hendon}. In fact, all-inorganic caesium lead iodide perovskite \ce{CsPbI3} has been reported to significantly increase device stability with comparable PV performance to the organic-inorganic hybrid PSCs~\cite{Hoffman,Kulbak,Kovalsky}. Actually, the smaller ionic radius of \ce{Cs+} cation, compared with \ce{MA+} cation, leads to an octahedral tilting and accordingly widening of band gap $E_\text{g}$ from 1.50 eV in \ce{MAPbI3}~\cite{Baikie} to 1.73 eV in \ce{CsPbI3} with a cubic phase~\cite{Eperon}, which is within the optimum range 1.7 $< E_\text{g} <$ 1.8 eV for a top cell material in tandem with a crystalline Si bottom cell~\cite{Ahmad}. It should be noted that other caesium halide perovskites such as \ce{CsPbBr3} and \ce{CsPbI3} have much higher band gaps of 2.3 eV and 3.0 eV, respectively~\cite{Kulbak,Heidrich}. Minding that crystalline \ce{CsPbI3} has two phases, {\it i.e.}, the photoactive black-phase with a cubic lattice ($\alpha$-\ce{CsPbI3}) at high temperature over 310 $^\circ$C  and photoinactive yellow-phase with an orthorhombic lattice ($\delta$-\ce{CsPbI3}) at lower temperatures~\cite{Eperon,Moller}, however, it is challenging to form the black-phase at room temperature, except the works of Snaith's group~\cite{Eperon} and Hoffman {\it et al}~\cite{Hoffman}. It is worth noting that nanoscale phases, such as quasi 2D nanoplate~\cite{Bekenstein,Swarnkar} and colloidal nanocrystal films~\cite{Davis,Akkerman,Nedelcu,Gomez,QLiu,Protesescu}, stabilize the cubic phase.

Alternatively, mixing the organic molecular cations with the inorganic alkali cations, forming double or triple mixed cations, has been reported to give rise to substantial enhancements of PV performance and phase stability for both single junction and multi-junction tandem solar cells~\cite{ZLi,Niu,Lee,McMeekin,Rehman,Saliba,Zhang,Duong,Saliba2}. Niu {\it et al.}~\cite{Niu} demonstrated that doping a small amount of Cs ($\sim$9\%) into \ce{MAPbI3} can stabilize the black perovskite structure, resulting in better thermal stability and device performance such as 18.1\% PCE. When replacing 10\% of FA with Cs in \ce{FAPbI3}, the PCE became as high as 19.0\%, which was attributed to the strengthened interaction between FA and I due to a contraction of cubo-octahedral volume by Cs doping~\cite{Lee}. Furthermore, Cs$-$FA$-$MA triple cation perovskites~\cite{Saliba} and mixed-cation, mixed-halide perovskites (Cs$-$FA, I$-$Br)~\cite{Rehman} have been also reported to exhibit better photo- and moisture-stability and moreover effective reduction of the crystallization temperature during the annealing process. On the other hand, it was established that adding Rb can achieve more stable perovskite by better tuning the Goldschmidt tolerance factor~\cite{Zhang,Duong,Saliba2}. Zhang {\it et al.} showed that 5\% Rb amount in Rb$-$FA$-$MA triple cation perovskite can provide the best film quality and the PCE of 20\% on a 65 mm$^2$ device~\cite{Zhang}. Very recently, it has been reported that Rb$-$Cs$-$MA$-$FA quadruple cation perovskites exhibit a band gap of 1.73 eV, negligible hysteresis, steady state efficiency as high as 17.4\%, and 26.4\% in tandem cell using a 23.9\% c-Si cell, close to the current record for a single junction Si cell of 26.6\%~\cite{Duong}.

These plenty of experimental findings require to systematically investigate the electronic structures, optical properties, and intrinsic material stability of \ce{RbPbI3} and \ce{CsPbI3} for a better insight of materials engineering. To the best of our knowledge, however, computational works on these compounds are scarce when compared with the extensive studies on \ce{MAPbI3} or \ce{FAPbI3}. Only one paper on the electronic band structures and dielectric properties of \ce{RbPbI3} with both $\alpha$- and $\delta$-phases has been published~\cite{Brgoch}, while there exist several works on \ce{CsPbI3}~\cite{Brgoch,Hendon,Murtaza,Afsari} and \ce{CsSnX3}~\cite{RYang,GSong,Silva,Huang}.

In this work, we perform systematic first-principles calculations of the all-inorganic iodide perovskites \ce{RbPbI3}, \ce{CsPbI3}, and their solid solutions \ce{Rb_{$1-x$}Cs_{$x$}PbI3} by using the virtual crystal approximation (VCA) method within density functional theory (DFT). Here, we regard that mixing Cs with Rb can stabilize the perovskite black-phase in addition to beneficial effect on band gap engineering, based on the insight that \ce{RbPbI3} does not undergo a phase transformation when elevating temperature~\cite{Brgoch}. Firstly we check the reliability of exchange-correlation functional with calculations of lattice constants and band gaps of $\alpha$-\ce{CsPbI3} and \ce{RbPbI3}. As increasing the Cs content $x$ from 0.0 to 1.0, we calculate the lattice constants, electronic band structures and density of states (DOS), and optical properties with the inclusion of van der Waals (vdW) dispersion terms. We performed additional calculations of DOS including the many-body interactions and spin-orbit coupling (SOC) effect with the GW+SOC method.

\section{\label{method}Computational Methods}
\begin{figure*}[!t]
\begin{center}
\begin{tabular}{l@{\hspace{40pt}}l}
(a) & (b)\\
\includegraphics[clip=true,scale=0.2]{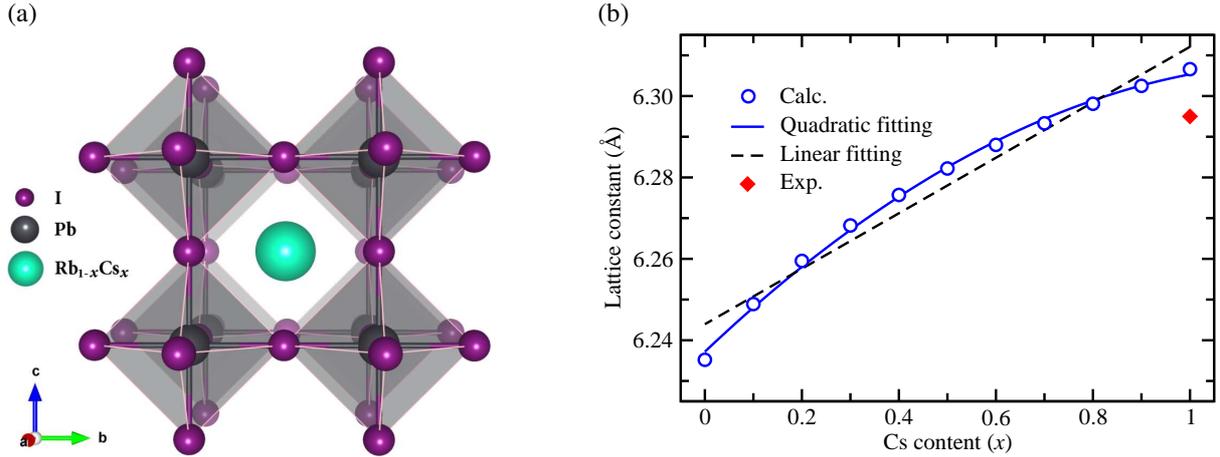} &
\includegraphics[clip=true,scale=0.5]{fig1b.eps} \\
\end{tabular}
\end{center}
\caption{\label{fig_lattice}(a) Polyhedral view of inorganic perovskite solid-solution \ce{Rb_{$1-x$}Cs_{$x$}PbI3}, and (b) their lattice constants as a function of Cs content $x$. The fourth polynomial fitting as well as linear fitting and experimental value for \ce{CsPbI3}~\cite{Trots} are shown.}
\end{figure*}

Most of the calculations were carried out by using the pseudopotential plane-wave method as implemented in Quantum ESPRESSO package~\cite{QE} (version 6.2). The ultrasoft pseudopotentials were generated by using the Atomic code provided in the package, where the valence electronic configurations of atoms are Cs-$6s^1$, Rb-$5s^1$, I-$5s^25p^5$, and Pb-$6s^26p^2$. As increasing the Cs content $x$ from 0.0 to 1.0 with a step of 0.1, we also generated pseudopotentials of virtual atoms \ce{Rb_{$1-x$}Cs_{$x$}} using the VCA method~\cite{yucj07}, as our previous studies for the hybrid mixed halide perovskites confirmed its efficiency and accuracy~\cite{yucj10,yucj12}. The plane-wave cutoff energy of 80 Ry and $k$-point mesh of (8$\times$8$\times$8) for structural optimization guaranteed a total energy convergence of 1 meV per unit cell. We obtained the equilibrium lattice constants by calculating the total energies at different unit cell volumes and then fitting the obtained $E-V$ data into the Birch-Murnaghan equation of states~\cite{birch}. All of the atomic positions were relaxed until the atomic forces were less than 0.02 eV \AA$^{-1}$. For better calculations of the electronic and optical properties, we used a denser $k$-point mesh of (12$\times$12$\times$12). The crystalline phases were supposed to be $\alpha$-phase with a $Pm$\={3}$m$ space group~\cite{Trots,Moller} as depicted in Fig.~\ref{fig_lattice}(a), since this phase is responsible for PSC performance and can be stabilized at room temperature~\cite{Eperon,Hoffman}.

The frequency-dependent dielectric constants, $\varepsilon(\omega)=\varepsilon_1(\omega)+i\varepsilon_2(\omega)$, were calculated within the density functional perturbation theory (DFPT)~\cite{Sharma} as implemented in the package. Then, the photoabsorption coefficients were obtained following Eq.~\ref{absorption}~\cite{yucj10,yucj12,yucj06},
\begin{equation}
\label{absorption}
\alpha(\omega)=\frac{2\omega}{c}\sqrt{\frac{\sqrt{\varepsilon^2_1(\omega)+\varepsilon^2_2(\omega)}-\varepsilon_1(\omega)}{2}},
\end{equation}
where $c$ is the velocity of light. Within the Mott-Wannier model, the exciton binding energies were calculated following Eq.~\ref{eig_exciton}
\begin{equation}
\label{eig_exciton}
E_b^\text{ex}=\frac{m_ee^4}{2(4\pi\varepsilon_0)^2\hbar^2}\frac{m_r^*}{m_e}\frac{1}{\varepsilon_s^2} \approx13.56\frac{m_r^*}{m_e}\frac{1}{\varepsilon_s^2}~(\text{eV}),
\end{equation}
where $m_r^*$ is the reduced effective mass evaluated by $1/m_r^*=1/m_e^*+1/m_h^*$ with the effective masses of electron $m_e^*$ and hole $m_h^*$, and $\varepsilon_s$ is the static dielectric constant, respectively.

To determine band gaps of \ce{RbPbI3} and \ce{CsPbI3} more precisely, we performed self-consistent GW calculations with a consideration of SOC effect, {\it i.e.}, GW+SOC calculations, using the VASP code~\cite{vasp1}. For these calculations, we used the cutoff energy of 500 eV for plane-wave basis set, $k$-point mesh of (4$\times$4$\times$4), and projector augmented wave (PAW) potentials, of which valence electronic configurations are Cs-$5s^25p^66s^1$, Rb-$4s^24p^65s^1$, I-$5s^25p^5$, and Pb-$6s^26p^2$. For the calculations of response function, smaller cutoff energy of 250 eV was used, together with the 80 Kohn-Sham (KS) states, where the first 34 of them were fully occupied and determined in the self-consistent field (SCF) calculations.

\section{Results and discussion}
\subsection{Assessment of exchange-correlation functional}
It is known that for inorganic compounds the exchange-correlation functionals within the local density approximation (LDA) overestimate a binding between atoms and underestimate a band gap by half of experimental value, whereas those within the generalized gradient approximation (GGA) improve the accuracy with slight underestimations of both binding and band gap. For example, Perdew-Burke-Ernzerhof (PBE) functional~\cite{pbe}, typical GGA functional, gave the lattice constant of $\alpha$-\ce{CsPbI3} as 6.39 \AA~versus 6.29 \AA~in experiment~\cite{Trots} (relative error 1.6\%) and its band gap as 0.60 eV versus 1.73 eV in experiment~\cite{Brgoch}. It should be noted that, for the organic-inorganic hybrid halide perovskite \ce{MAPbI3}, PBE functional can almost exactly reproduce the experimental band gap due to the fortuitous error cancellation between the underestimation by PBE and overestimation by ignoring SOC effect~\cite{Even13}, and the addition of vdW dispersion terms also slightly improves both the lattice constant and band gap~\cite{yucj06,yucj10,yucj12}. Here, we tested the reliability of PBE, PBE revised version for solid (PBEsol)~\cite{PBEsol}, and PBE augmented by vdW interaction (PBE+vdW)~\cite{vdwDF2} with lattice constant and band gap.

Table~\ref{tab_lattice} shows the calculated lattice constants and band gaps of $\alpha$-\ce{CsPbI3} and $\alpha$-\ce{RbPbI3} with the available experimental values and previously calculated values for comparison. The lattice constants were determined by the procedure described in the Method section~\ref{method}, {\it i.e.}, by varying unit cell volumes evenly from 0.9$V_0$ to 1.1$V_0$ with the pre-determined equilibrium volume $V_0$, calculating their total energies versus volume, and fitting the obtained $E-V$ data into the Birch-Murnaghan equation of states~\cite{birch}. It is found from Table~\ref{tab_lattice} that for \ce{CsPbI3}, PBE overestimates the experimental lattice constant with a relative error of 1.3\% and PBEsol underestimates with $-$0.8\% in accordance with the previous calculations~\cite{Hendon,Brgoch}, whereas PBE+vdW gives the closest value with a slight overestimation of 0.5\%. This indicates that the addition of vdW interaction terms still improves the accuracy of calculations for all-inorganic halide perovskites as pointed out by Brgoch {\it et al}~\cite{Brgoch}. It should be noted that for $\alpha$-\ce{RbPbI3} experimental value is not available and in Ref.~\cite{Brgoch} they strangely presented the same value to $\alpha$-\ce{CsPbI3}, while in our calculation its lattice constant was calculated to be smaller by 0.06$-$0.09 \AA~than \ce{CsPbI3} with a rational soundness due to the smaller ionic radius of \ce{Rb+} than \ce{Cs+}.

\begin{table}[!b]
\caption{\label{tab_lattice}Lattice constant ($a$) and band gap ($E_\text{g}$) of inorganic perovskites \ce{CsPbI3} and \ce{RbPbI3}, calculated using different exchange-correlation functionals. Experimental and previously calculated values are listed.}
\small
\begin{tabular}{l@{\hspace{5pt}}l@{\hspace{5pt}}cccc@{\hspace{0pt}}cc}
\hline
 & & \multicolumn{3}{c}{This work} && \multicolumn{2}{c}{Previous work} \\
\cline{3-5} \cline{7-8}
           &           &  PBE  & PBEsol & PBE+vdW&&   Exp.   &   Calc. \\ 
\hline
\ce{CsPbI3}&$a$ (\AA)  & 6.37  &  6.24  &  6.32  && 6.29$^a$ & 6.26$^c$ \\ 
           &           &       &        &        &&          & 6.39$^d$ \\
           &           &       &        &        &&          & 6.18$^e$ \\
           &$E_\text{g}$ (eV) & 1.48  &  1.35  &  1.50  && 1.73$^b$ & 0.51$^c$ \\
           &           &       &        &        &&          & 0.60$^d$ \\
           &           &       &        &        &&          & 1.30$^e$ \\
\hline
\ce{RbPbI3}&$a$ (\AA)  & 6.31  &  6.18  &  6.23  &&    -     & 6.39$^d$ \\
           &$E_\text{g}$ (eV) & 1.45  &  1.26  &  1.40  &&    -     & 0.73$^d$ \\
\hline
\end{tabular}
$^a$~Ref.~\cite{Trots}, $^b$~Ref.~\cite{Eperon} \\
$^c$~PBEsol for lattice constant and HSE06+SOC for band gap~\cite{Hendon} \\
$^d$~PBE for lattice constant and HSE06+SOC for band gap~\cite{Brgoch} \\
$^e$~FP-LAPW with Wu-Cohen GGA functional~\cite{Murtaza}
\end{table}

\begin{figure*}[!th]
\begin{center}
\begin{tabular}{l@{\hspace{1pt}}ll}
~~~~(a) & \multicolumn{2}{l}{~~~~~~~~~~~(c)} \\
~~~~\includegraphics[clip=true,scale=0.47]{fig2a.eps} & \multicolumn{2}{c}{\includegraphics[clip=true,scale=0.45]{fig2c.eps}}\\
& & \\
(b)& (d) & (e) \\
\includegraphics[clip=true,scale=0.45]{fig2b.eps} &
\includegraphics[clip=true,scale=0.21]{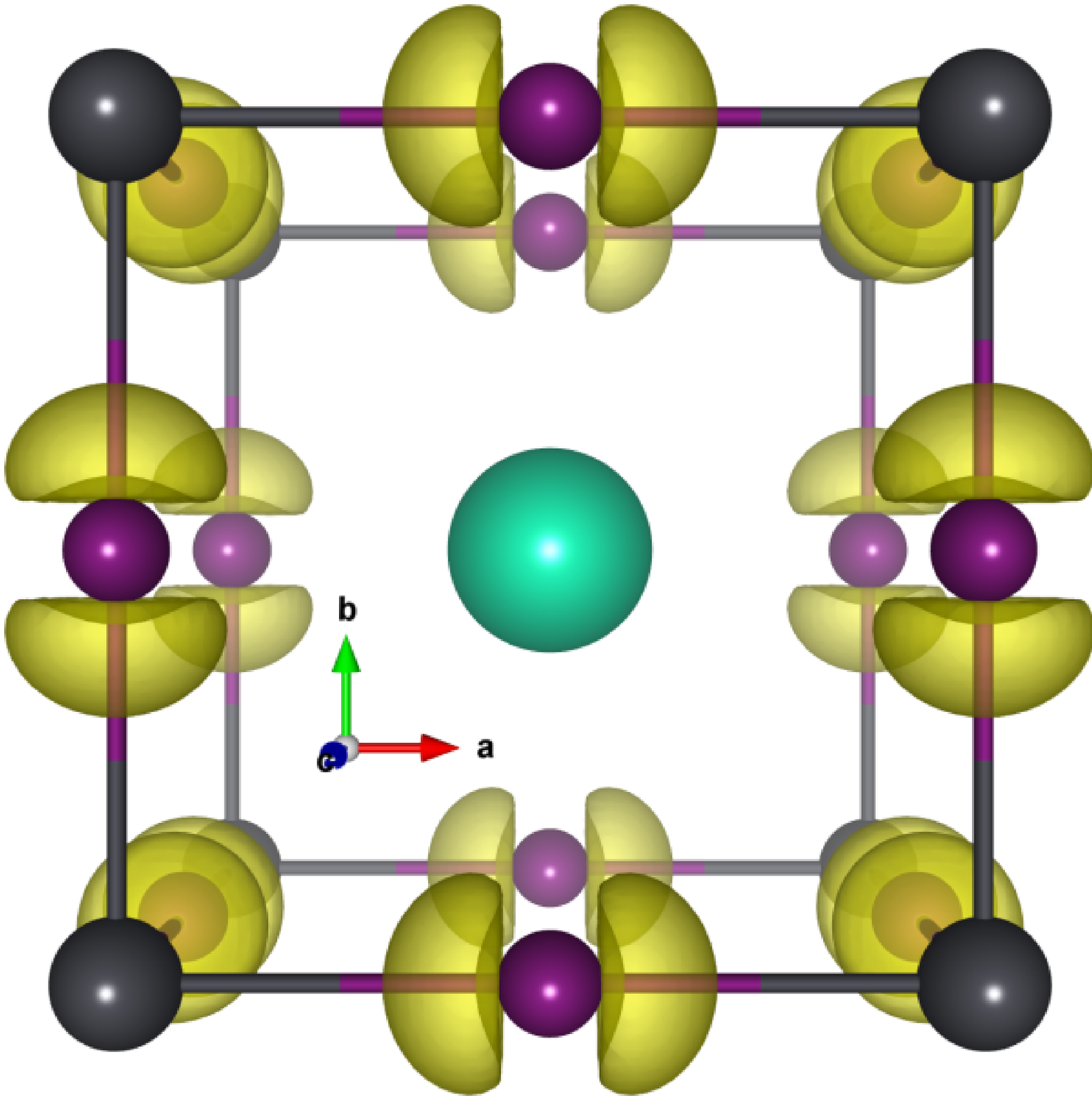} &
\includegraphics[clip=true,scale=0.21]{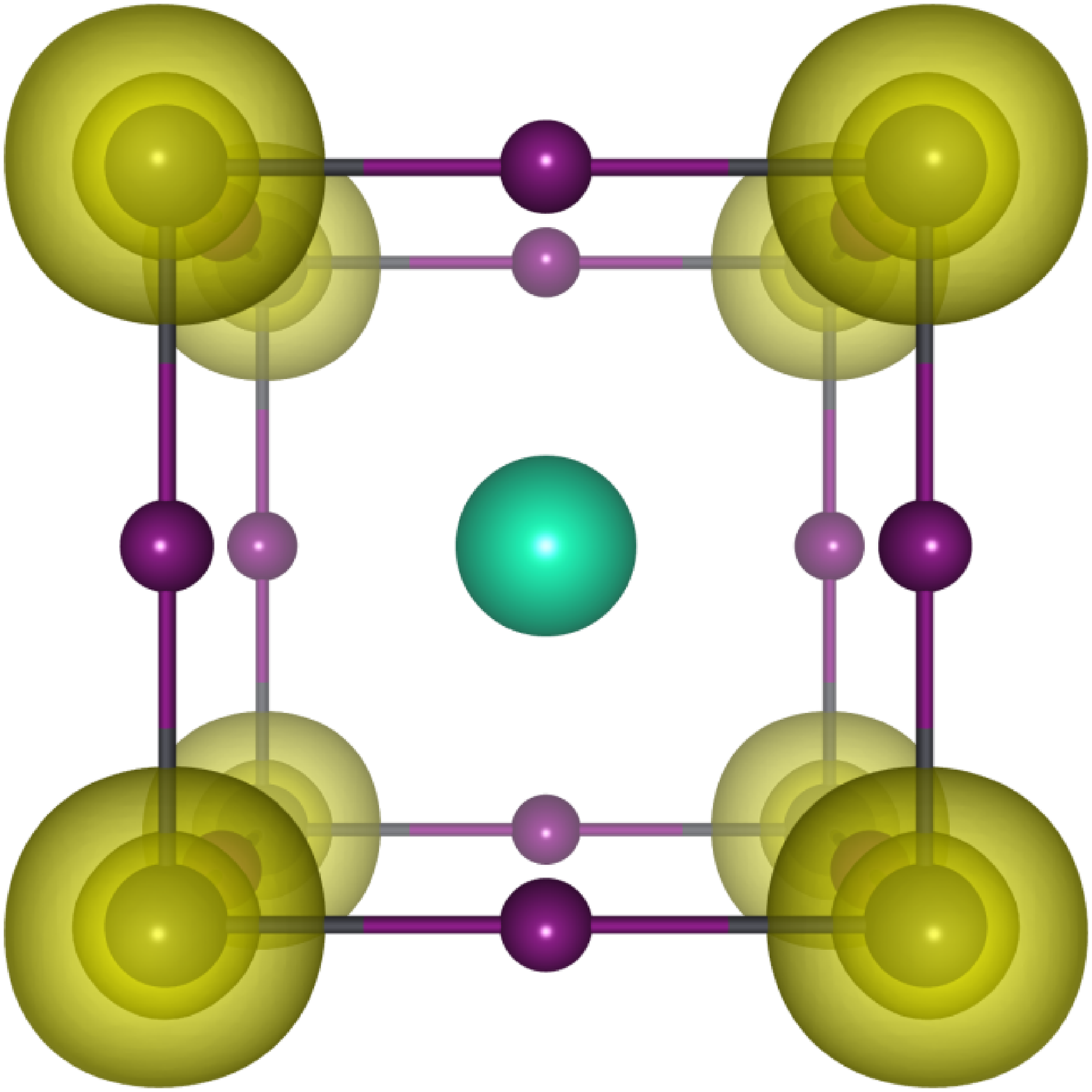} \\
\end{tabular}
\end{center}
\caption{\label{fig_electronic}Systematic change in electronic properties of inorganic iodide perovskite solid-solutions \ce{Rb_{$1-x$}Cs_{$x$}PbI3} as increasing the Cs content $x$, calculated with PBE+vdW method. (a) Electronic band structures with (dashed line) or without (solid line) SOC effect, and (b) band gaps as a function of Cs content $x$, with the fourth polynomial as well as linear fitting lines. (c) Total and atomic resolved electronic density of states, setting the valence band maximum (VBM) to be zero. The charge density isosurface around (d) VBM and (e) conduction band minimum of \ce{CsPbI3} at the value of 0.05~$|e|$ \AA$^{-3}$.}
\end{figure*}

Many of previous DFT investigations~\cite{Umari,Even13} has verified an important role of relativistic effect from the heavy metallic elements in the hybrid halide perovskites, which can be considered by scalar relativistic effect approximately or by SOC effect in higher level. In order to check the importance of incorporating relativistic effects in the case of inorganic halide perovskites, we calculated electronic band structures with or without SOC contribution as shown in Fig.~\ref{fig_electronic}(a). Table~\ref{tab_lattice} presents only band gaps calculated without SOC effects. Interestingly, for $\alpha$-\ce{CsPbI3}, the calculated band gaps without SOC effects were in reasonable agreement with the experimental value of 1.73 eV~\cite{Eperon} with slight underestimations of 0.25 eV (PBE), 0.38 eV (PBEsol) and 0.23 eV (PBE+vdW), again indicating the best agreement by the PBE+vdW functional. When including the SOC contribution, the deviation of band gap from the experimental value became 1.20 eV, which is comparable with those in the case of hybrid halide perovskite. As clearly shown in Fig.~\ref{fig_electronic}(a), this is due to a split of the degenerate, unoccupied \ce{Pb2+} $p$-orbitals and their down-shift by the SOC effect. Such large deviations over 1 eV were also observed in the previous computational studies~\cite{Hendon,Brgoch}, where HSE06+SOC produced the band gaps of 0.51$\sim$0.60 eV, resulting in the deviations of 1.22$\sim$1.13 eV. Therefore, it became obvious that GGA or GGA+vdW for better without SOC effect are likely to describe the structural and electronic properties of inorganic halide perovskites as well as hybrid ones. It is worth noting that many-body interaction coupled with SOC effect, {\it i.e.}, GW+SOC, can give the most reliable band gaps, as we present later in this work. Henceforth, we progressed the work with PBE+vdW without the SOC effect, aiming at finding the variation tendency of properties of their solid solutions \ce{Rb_{$1-x$}Cs_{$x$}PbI3} as increasing $x$.

\subsection{Lattice constants and electronic structures}

\begin{figure*}[!th]
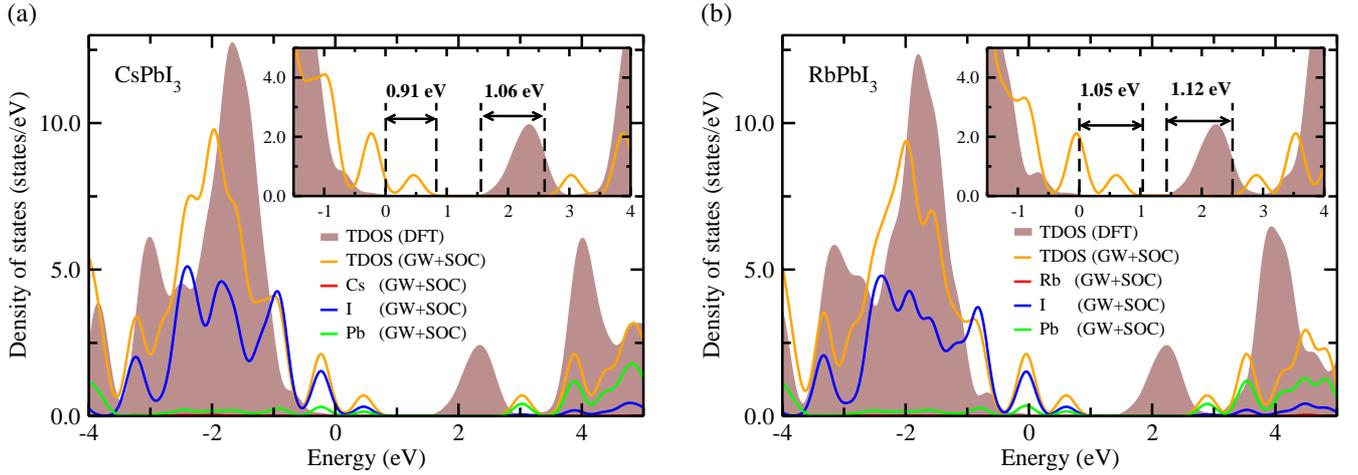

\begin{center}
\begin{tabular}{l@{\hspace{20pt}}l}
(a)&(b)\\
\includegraphics[clip=true,scale=0.52]{fig3a.eps} &
\includegraphics[clip=true,scale=0.52]{fig3b.eps}\\
\end{tabular}
\end{center}
\caption{\label{fig_gw}Total and atomic resolved electronic density of states (DOS) of (a) \ce{CsPbI3} and (b) \ce{RbPbI3}, calculated by using GW+SOC method. For a comparison, TDOS calculated by using the PBE+vdW functional is also shown. Insets present the shifts of the VBM and CBM explicitly.}
\end{figure*}

Fig.~\ref{fig_lattice}(b) shows the lattice constants of inorganic iodide perovskite solid-solutions \ce{Rb_{$1-x$}Cs_{$x$}PbI3} with the quadratic and linear fitting lines. It is shown that as increasing the Cs content $x$, the lattice constant increases. Unlike the hybrid mixed halide perovskite \ce{MAPb(I_{$1-x$}Br_{$x$})3}~\cite{yucj10} or \ce{MAPb(I_{$1-x$}Cl_{$x$})3}~\cite{yucj12}, however, the lattice constants slightly deviate from the Vegard's law, {\it i.e.}, their increasing tendency does not follow the linear function of Cs content $x$. That is why we tried to perform the quadratic fitting, resulting in a function of $a(x)=6.237+0.113x-0.045x^2$ (\AA). It should be noted that the difference of lattice constants between \ce{RbPbI3} and \ce{CsPbI3} is not large (0.09 \AA) compared with the case of mixed halide perovskites (0.3$-$0.7 \AA). Since the deviation is not far from the linear fitting, we also tried to obtain the linear function as $a(x)=6.244+0.068x$ (\AA).

We show systematic change in electronic band structures of \ce{Rb_{$1-x$}Cs_{$x$}PbI3} along the edge line of X$-$R$-$X$_1$ in reciprocal space (see Fig.S1 for the entire bands) as gradually increasing the Cs content $x$ in Fig.~\ref{fig_electronic}(a), where the valence band maximum (VBM) is set to be zero. It was found that for all Cs contents, the PBE+vdW functional without SOC contribution yields the valence bands (VBs; blue solid lines) almost coincident with those by incorporating SOC effect (blue dashed lines), while the conduction bands (CBs) degenerated at R point (red solid lines) compared with those splitted and down-shifted by SOC effect (red dashed lines). This resulted in the decrease of band gaps as much as $\sim$1.2 eV when including the SOC effect. It should be noted that the band gaps are characterised by a direct mode at R point of reciprocal space, being similar to the hybrid halide perovskites. This is one of the best advantages for a good solar absorber, since the charge carriers such as electron and hole could be directly generated by absorption of photons without any energy loss by interacting with phonons.

Like lattice constants, the band gap $E_\text{g}$ also increases going from \ce{RbPbI3} to \ce{CsPbI3} as shown in Fig.~\ref{fig_electronic}(b). Actually, when replacing the \ce{Cs+} cation in $\alpha$-\ce{APbI3} by the \ce{Rb+} cation with the smaller ionic radius, the interaction between Pb and I atoms within \ce{PbI6} octahedra can be enhanced, resulting in the contraction of crystalline lattice and accordingly increase of band gap. It was known that the variation tendency of band gaps as increasing the concentration of composition in the hybrid mixed halide perovskites can be described by a quadratic function~\cite{Atourki,yucj10,yucj12}. Interestingly, the band gap in the Rb$-$Cs inorganic halide perovskites also follows the quadratic function of Cs content $x$ as $E_\text{g}(x)=1.407+0.164x-0.071x^2$ (eV), which can be reduced into the following equation,
\begin{equation}
\label{eq_band}
E_\text{g}(x)=E_\text{g}(0)+[E_\text{g}(1)-E_\text{g}(0)-b]x+bx^2,
\end{equation}
where $E_\text{g}(0)$ = 1.407 eV is the band gap of \ce{RbPbI3}, $E_\text{g}(1)$ = 1.500 eV is the band gap of \ce{CsPbI3}, and $b=-$0.071 eV is the bowing parameter, reflecting the fluctuation degree in the crystal field. The first notable difference from the hybrid mixed halide perovskites is the sign of the bowing parameter, which is minus in the case of inorganic mixed alkali iodide perovskites versus plus in the case of hybrid mixed halide perovskites. Since the bowing parameter is quite small, compositional disorder can be said to be low and miscibility between \ce{RbPbI3} and \ce{CsPbI3} to be good. Moreover, the decrease of band gap as increasing the Rb content indicates an improvement of light absorption property by mixing Cs with Rb.

In order to find out the role of electrons of constituent atoms in the band structure, the electronic density of states (DOS) were calculated and presented in Fig.~\ref{fig_electronic}(c) (see Fig.S2 for partial DOS). We see that the upper VBs and lower CBs responsible for transport of charge carriers (electron and hole) come from the overlap of I 5$p$/Pb 6$s$ states and the Pb 6$p$ state, respectively. This is rational because the \ce{Pb2+} cation has a filled 6$s$ state and an empty 6$p$ state, while the \ce{I-} anion has a completely filled 5$p$ state. The 5/6$s$ atomic orbitals of Rb/Cs are found far below and above from the VBM and conduction band minimum (CBM), indicating that \ce{(Rb_{$1-x$}Cs_{$x$})+} (x=[0, 1]) cation is electronically inert~\cite{Hendon}. Similar observations were found on the hybrid mixed halide perovskites~\cite{yucj10,yucj12}. Notably, it was observed that the VBM is characterised by the I 5$p$ state and the CBM by the Pb 6$s$ state, as shown in Fig.~\ref{fig_electronic}(d) and (e) (see Fig.S3 for those of \ce{RbPbI3}), whereas in the case of hybrid halide perovskites the VBM is formed from the overlap of I 5$p$ and Pb 6$s$ states and the CBM from the Pb 6$p$ state.

\begin{figure*}[!th]
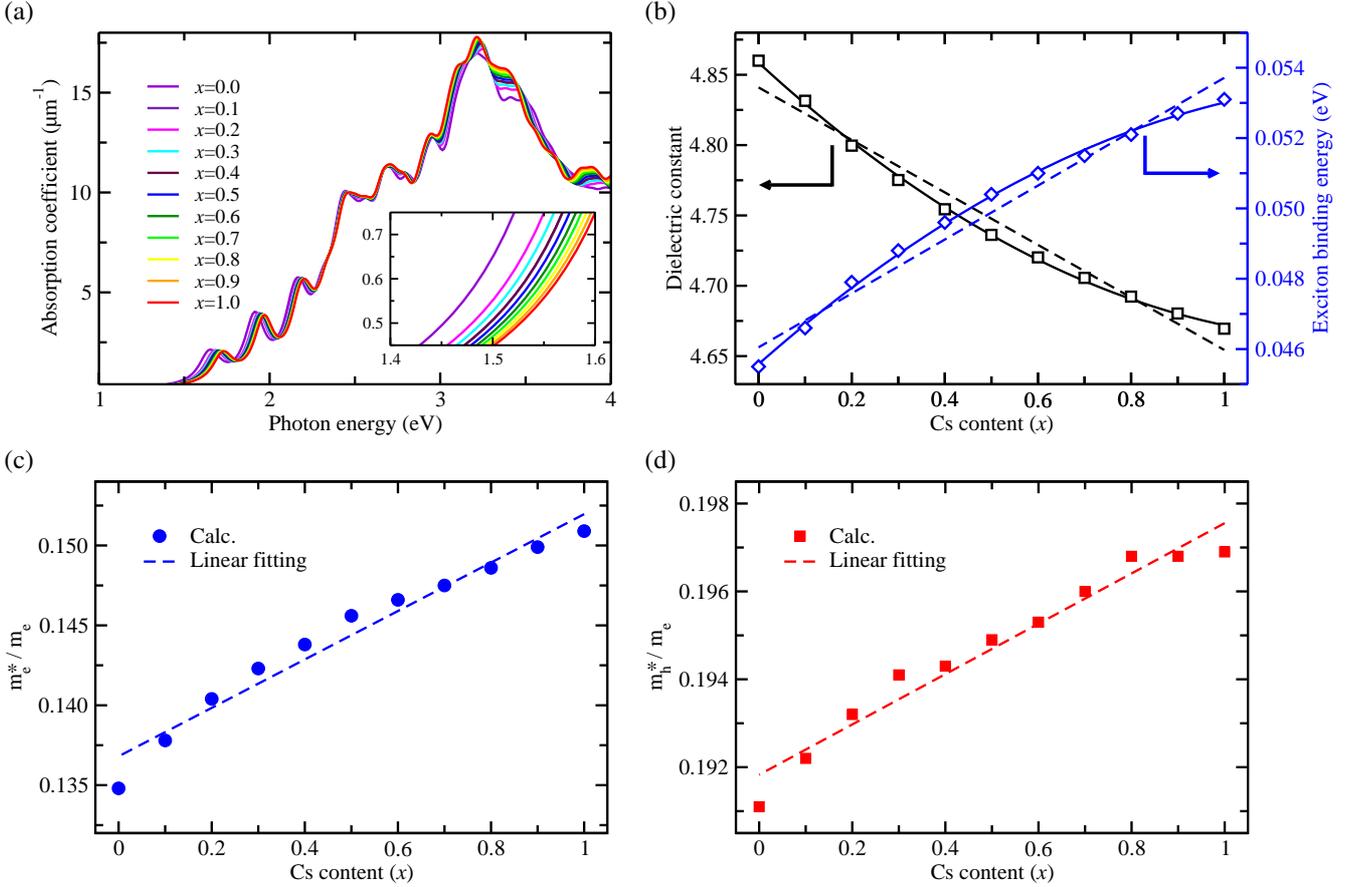

\begin{center}
\begin{tabular}{l@{\hspace{10pt}}l}
(a)&(b)\\
~~~~\includegraphics[clip=true,scale=0.48]{fig4a.eps}&
~~~\includegraphics[clip=true,scale=0.48]{fig4b.eps}\\
(c)&(d)\\
\includegraphics[clip=true,scale=0.48]{fig4c.eps} &
\includegraphics[clip=true,scale=0.48]{fig4d.eps}\\
\end{tabular}
\end{center}
\caption{\label{fig_absorption}(a) Photoabsorption coefficients, (b) static dielectric constants and exciton binding energies, and (c) and (d) effective masses of electron and hole in the inorganic iodide perovskite solid-solutions \ce{Rb_{$1-x$}Cs_{$x$}PbI3} as increasing the Cs content $x$ from 0 to 1, calculated with PBE+vdW method.}
\end{figure*}

As we discussed above, the band gaps were remarkably underestimated by including the SOC effect due to the downshift of CBM composed of Pb atomic orbitals. This can be cured by adding the effect of many-body interactions within the GW approximation. We only treated the pure \ce{RbPbI3} and \ce{CsPbI3} compounds with the optimized structures by using the PBE+vdW functional and considering the SOC effect, to see the improvement of band gap calculation with the GW+SOC method. Provided that the number of valence electrons in the unit cell is 34, which can be determined from the valence electronic configurations of PAW potentials (see Method section~\ref{method}) for the constituent atoms, the occupied Kohn-Sham (KS) states of 17 were considered in this calculation. Fig.~\ref{fig_gw} shows their DOSs calculated by GW+SOC method, together with TDOS calculated by using the PBE+vdW functional for a comparison. It was found that for \ce{CsPbI3} both the DFT VBM and CBM are pushed up by 0.91 and 1.06 eV, respectively, resulting in a totally 0.15 eV upshift of the DFT band gap (1.50 eV) and thus the improved band gap of 1.65 eV with only a deviation of 0.08 eV from the experimental value of 1.73 eV. For the case of \ce{RbPbI3}, the upshift of VBM and CBM are 1.05 and 1.12 eV, respectively, which gives a total upshift of the DFT band gap (1.40 eV) by 0.07 eV.

\subsection{Optical properties and charge carrier transport}
The photoabsorption coefficient is major optical property that determines the capability of light absorber for harvesting solar energy. This can be obtained by using Eq.~\ref{absorption} with the calculated macroscopic dielectric constants as a function of frequency. Our previous calculations for the hybrid halide perovskites~\cite{yucj12} gave an instruction that the DFPT approach, considering the effect of atomic displacements, can yield the almost identical dielectric constants with those from the Bethe-Salpeter approach including the excitonic and many-body effects. In this work, therefore, we performed the DFPT calculations of the macroscopic dielectric constants with their real and imaginary parts (see Fig.S4). In Fig.~\ref{fig_absorption}(a), we plotted the photoabsorption coefficients as a function of photon energy (frequency) as increasing the Cs content $x$. For higher Cs contents, since the band gap gradually increases as discussed above, the absorption onsets are shifted, at a slow pace, to the ultraviolet region in solar spectrum, indicating a harshening of photoabsorption capability. The maximum photoabsorption coefficients were found to be slightly lower as $\sim$17 $\mu\text{m}^{-1}$ at the position of $\sim$3.25 eV, compared with those ($\sim$22 $\mu\text{m}^{-1}$) of the hybrid halide perovskites~\cite{yucj10,yucj12}.

The static dielectric constants were extracted from the frequency dependent macroscopic dielectric functions at the zero photon energy to be used for the calculation of exciton binding energy. As shown in Fig.~\ref{fig_absorption}(b), they decrease in accordance with the quadratic function of Cs content $x$ like $\varepsilon_\text{s}(x)=4.859-0.304x+0.117x^2$, being different from the linear function in the case of the hybrid halide perovskites~\cite{yucj12}. We also tried to fit them into the linear function, yielding $\varepsilon_\text{s}(x)=4.841-0.187x$. For pure \ce{CsPbI3} (\ce{RbPbI3}), our calculated value of 4.7 (4.9) is in reasonable agreement with the previous theoretical value of 5.3 (4.8) calculated with the PBE+SOC approach~\cite{Brgoch}. On the other hand, these values are comparable with those of the hybrid halide perovskite, {\it e.g.}, 5.6 of \ce{MAPbI3}~\cite{yucj12}.

In order to see how the charge carriers such as electron and hole behave after being generated by absorbing photons, we calculated effective masses of electron ($m_e^*$) and hole ($m_h^*$), and exciton binding energy, which are useful for a qualitative estimation of charge carrier mobility. The effective masses of electron and hole were calculated around the CBM and VBM at the R point using the refined energy band structure within the parabolic approximation. As shown in Fig.~\ref{fig_absorption}(c) and (d), they slightly increase along the linear functions of Cs content as $m_e^*(x)=(0.137+0.015x)m_e$ and $m_h^*(x)=(0.192+0.006x)m_e$ respectively, indicating an enhancement of charge carrier mobilities upon the Rb addition to \ce{CsPbI3}. For pristine \ce{CsPbI3}, the calculated values of $m_e^* = 0.15~m_e$ and $m_h^* = 0.20~m_e$ are comparable with the previous theoretical values of $m_e^* = 0.22~m_e$ and $m_h^* = 0.19~m_e$ with the modified Becke-Johnson approach~\cite{Afsari} and $m_h^* = 0.16~m_e$ with the HSE06+SOC approach~\cite{Hendon}. Moreover, these are slightly lower compared with those of \ce{MAPbI3} obtained with the same level of theory~\cite{yucj12} ($m_e^* = 0.20~m_e$ and $m_h^* = 0.23~m_e$), indicating lighter carriers in \ce{CsPbI3}, which might be very promising for the efficiency of solar cells.

Finally, we considered the exciton binding energies $E_\text{b}$ calculated by Eq.~\ref{eig_exciton}, using the static dielectric constants and charge carrier effective masses. Since the exciton binding energy is a direct estimation for the electrostatic attraction between electron and hole, its smaller value indicates free-like charge carriers. As shown in Fig.~\ref{fig_absorption} (b), they increase according to the quadratic function of $E_\text{b}(x)=0.046+0.012x-0.004x^2$ (eV) as increasing the Cs content, due to the decrease of the static dielectric constants and increase of the charge carrier effective masses. This again indicates an enhancement of solar cell performance upon the Rb addition to \ce{CsPbI3}. Furthermore, these are comparable to \ce{MAPbI3} and the inorganic thin-film semiconductors, indicating that the charge carriers in the all-inorganic alkali iodide perovskite, generated by photoabsorption, can behave as if they are free.

\section{Conclusions}
Using the virtual crystal approximation method, we have systematically investigated the electronic and optical properties of the inorganic alkali iodide perovskite solid-solutions \ce{Rb_{$1-x$}Cs_{$x$}PbI3}. Through the assessment of exchange-correlation functional, the PBE+vdW approach was selected to be the most reliable for both lattice constant and band gap. When increasing the Cs content $x$ from 0.0 to 1.0, the lattice constants and band gaps increase following the quadratic functions of $a(x)=6.237+0.113x-0.045x^2$ (\AA) and $E_\text{g}(x)=1.407+0.164x-0.071x^2$ eV, respectively, indicating a red-shift of absorption onset in \ce{RbPbI3} compared to \ce{CsPbI3}. It was found that the upper VBs and lower CBs come from the overlap of I 5$p$/Pb 6$s$ states and the Pb 6$p$ states. The GW+SOC approach was confirmed to improve the band gap calculation giving the band gap of 1.65 eV for \ce{CsPbI3} in good agreement with the experimental value of 1.73 eV, though the PBE+SOC approach gave the worst value. The photoabsorption coefficients were obtained using the real and imaginary parts of macroscopic dielectric functions calculated by the DFPT method. As increasing the Cs content, the static dielectric constants decrease following the quadratic function of $\varepsilon_\text{s}(x)=4.859-0.304x+0.117x^2$, and accordingly, the exciton binding energies increase according to the quadratic function of $E_\text{b}(x)=0.046+0.012x-0.004x^2$ (eV). Together with the linear decreases of electron and hole effective masses, these indicate a possible enhancement of solar cell performance upon the Rb addition to \ce{CsPbI3}.

\section*{Acknowledgments}
This work was supported partially by the State Committee of Science and Technology, Democratic People's Republic of Korea, under the state research project ``Design of Innovative Functional Materials for Energy and Environmental Application'' (No. 2016-20). The calculations have been carried out on the HP Blade System C7000 (HP BL460c) that is owned by Faculty of Materials Science, Kim Il Sung University. AW is supported by a Royal Society University Research Fellowship and the Leverhulme Trust.

\section*{Appendix A. Supplementary data}
Supplementary data related to this article can be found at URL.

\section*{Notes}
The authors declare no competing financial interest.

\bibliographystyle{elsarticle-num-names}
\bibliography{Reference}

\begin{thebibliography}{61}
\providecommand{\natexlab}[1]{#1}
\providecommand{\url}[1]{\texttt{#1}}
\providecommand{\urlprefix}{URL }
\expandafter\ifx\csname urlstyle\endcsname\relax
  \providecommand{\doi}[1]{doi:\discretionary{}{}{}#1}\else
  \providecommand{\doi}[1]{doi:\discretionary{}{}{}\begingroup
  \urlstyle{rm}\url{#1}\endgroup}\fi
\providecommand{\bibinfo}[2]{#2}

\bibitem[{Liu et~al.(2013)Liu, Johnston, and Snaith}]{Liu}
\bibinfo{author}{M.~Liu}, \bibinfo{author}{M.~B. Johnston},
  \bibinfo{author}{H.~J. Snaith}, \bibinfo{title}{Efﬁcient planar
  heterojunction perovskite solar cells by vapor deposition},
  \bibinfo{journal}{Nature} \bibinfo{volume}{501} (\bibinfo{year}{2013})
  \bibinfo{pages}{395--398}.

\bibitem[{Burschka et~al.(2013)Burschka, Pellet, Moon, Humphry-Baker, Gao,
  Nazeeruddin, and Gr\"{a}tzel}]{Burschka}
\bibinfo{author}{J.~Burschka}, \bibinfo{author}{N.~Pellet},
  \bibinfo{author}{S.~J. Moon}, \bibinfo{author}{R.~Humphry-Baker},
  \bibinfo{author}{P.~Gao}, \bibinfo{author}{M.~K. Nazeeruddin},
  \bibinfo{author}{M.~Gr\"{a}tzel}, \bibinfo{title}{Sequential deposition as a
  route to high-performance perovskite-sensitized solar cells},
  \bibinfo{journal}{Nature} \bibinfo{volume}{499} (\bibinfo{year}{2013})
  \bibinfo{pages}{316--319}.

\bibitem[{Yang et~al.(2017{\natexlab{a}})Yang, Park, Jung, Jeon, Kim, Lee,
  Shin, Seo, Kim, Noh, and Seok}]{WSYang}
\bibinfo{author}{W.~S. Yang}, \bibinfo{author}{B.~W. Park},
  \bibinfo{author}{E.~H. Jung}, \bibinfo{author}{N.~J. Jeon},
  \bibinfo{author}{Y.~C. Kim}, \bibinfo{author}{D.~U. Lee},
  \bibinfo{author}{S.~S. Shin}, \bibinfo{author}{J.~Seo},
  \bibinfo{author}{E.~K. Kim}, \bibinfo{author}{J.~H. Noh},
  \bibinfo{author}{S.~I. Seok}, \bibinfo{title}{{Iodide Management in
  Formamidinium-Lead-Halide-Based Perovskite Layers for Efficient Solar
  Cells}}, \bibinfo{journal}{Science} \bibinfo{volume}{356}
  (\bibinfo{year}{2017}{\natexlab{a}}) \bibinfo{pages}{1376--1379}.

\bibitem[{Ganose et~al.(2017)Ganose, Savory, and Scanlon}]{Ganose}
\bibinfo{author}{A.~M. Ganose}, \bibinfo{author}{C.~N. Savory},
  \bibinfo{author}{D.~O. Scanlon}, \bibinfo{title}{Beyond methylammonium lead
  iodide: prospects for the emergent field of ns$^2$ containing solar
  absorbers}, \bibinfo{journal}{Chem. Commun.} \bibinfo{volume}{53}
  (\bibinfo{year}{2017}) \bibinfo{pages}{20--44}.

\bibitem[{Baikie et~al.(2013)Baikie, Fang, Kadro, Schreyer, Wei, Mhaisalkar,
  Gr\"{a}tzel, and White}]{Baikie}
\bibinfo{author}{T.~Baikie}, \bibinfo{author}{Y.~N. Fang},
  \bibinfo{author}{J.~M. Kadro}, \bibinfo{author}{M.~Schreyer},
  \bibinfo{author}{F.~X. Wei}, \bibinfo{author}{S.~G. Mhaisalkar},
  \bibinfo{author}{M.~Gr\"{a}tzel}, \bibinfo{author}{T.~J. White},
  \bibinfo{title}{Synthesis and crystal chemistry of the hybrid perovskite
  (CH$_3$NH$_3$)PbI$_3$ for solid-state sensitised solar cell applications},
  \bibinfo{journal}{J. Mater. Chem. A} \bibinfo{volume}{1}
  (\bibinfo{year}{2013}) \bibinfo{pages}{5628--5641}.

\bibitem[{D'Innocenzo et~al.(2014)D'Innocenzo, Grancini, Alcocer, Kandada,
  Stranks, Lee, Lanzani, Snaith, and Petrozza}]{Innocenzo}
\bibinfo{author}{V.~D'Innocenzo}, \bibinfo{author}{G.~Grancini},
  \bibinfo{author}{M.~J.~P. Alcocer}, \bibinfo{author}{A.~R.~S. Kandada},
  \bibinfo{author}{S.~D. Stranks}, \bibinfo{author}{M.~M. Lee},
  \bibinfo{author}{G.~Lanzani}, \bibinfo{author}{H.~J. Snaith},
  \bibinfo{author}{A.~Petrozza}, \bibinfo{title}{Excitons versus free charges
  in organo-lead tri-halide perovskites}, \bibinfo{journal}{Nat. Commun.}
  \bibinfo{volume}{5} (\bibinfo{year}{2014}) \bibinfo{pages}{3486}.

\bibitem[{Shao et~al.(2014)Shao, Xiao, Bi, Yuan, and Huang}]{Shao}
\bibinfo{author}{Y.~Shao}, \bibinfo{author}{Z.~Xiao}, \bibinfo{author}{C.~Bi},
  \bibinfo{author}{Y.~Yuan}, \bibinfo{author}{J.~Huang}, \bibinfo{journal}{Nat.
  Commun.} \bibinfo{volume}{5} (\bibinfo{year}{2014}) \bibinfo{pages}{5784}.

\bibitem[{Walsh et~al.(2015)Walsh, Scanlon, Chen, Gong, and Wei}]{Walsh}
\bibinfo{author}{A.~Walsh}, \bibinfo{author}{D.~O. Scanlon},
  \bibinfo{author}{S.~Chen}, \bibinfo{author}{X.~G. Gong},
  \bibinfo{author}{S.-H. Wei}, \bibinfo{title}{Self-regulation mechanism for
  charged point defects in hybrid halide perovskites}, \bibinfo{journal}{Angew.
  Chem. Int. Ed.} \bibinfo{volume}{54} (\bibinfo{year}{2015})
  \bibinfo{pages}{1791--1794}.

\bibitem[{Li et~al.(2016{\natexlab{a}})Li, Li, Zheng, Gao, and Huang}]{Li}
\bibinfo{author}{B.~Li}, \bibinfo{author}{Y.~Li}, \bibinfo{author}{C.~Zheng},
  \bibinfo{author}{D.~Gao}, \bibinfo{author}{W.~Huang},
  \bibinfo{title}{Advancements in the stability of perovskite solar cells:
  degradation mechanisms and improvement approaches}, \bibinfo{journal}{RSC
  Adv.} \bibinfo{volume}{6} (\bibinfo{year}{2016}{\natexlab{a}})
  \bibinfo{pages}{38079--38091}.

\bibitem[{Berhe et~al.(2016)Berhe, Su, Chen, Pan, Cheng, Chen, Tsai, Chen,
  Dubaleb, and Hwang}]{Berhe}
\bibinfo{author}{T.~A. Berhe}, \bibinfo{author}{W.-N. Su},
  \bibinfo{author}{C.-H. Chen}, \bibinfo{author}{C.-J. Pan},
  \bibinfo{author}{J.-H. Cheng}, \bibinfo{author}{H.-M. Chen},
  \bibinfo{author}{M.-C. Tsai}, \bibinfo{author}{L.-Y. Chen},
  \bibinfo{author}{A.~A. Dubaleb}, \bibinfo{author}{B.-J. Hwang},
  \bibinfo{title}{Organometal halide perovskite solar cells: degradation and
  stability}, \bibinfo{journal}{Energy Environ. Sci.} \bibinfo{volume}{9}
  (\bibinfo{year}{2016}) \bibinfo{pages}{323--356}.

\bibitem[{Li et~al.(2017)Li, Cao, Yu, Chen, Sun, Shen, Zhu, Wang, Wei, Wu, and
  Zeng}]{XLi}
\bibinfo{author}{X.~Li}, \bibinfo{author}{F.~Cao}, \bibinfo{author}{D.~Yu},
  \bibinfo{author}{J.~Chen}, \bibinfo{author}{Z.~Sun},
  \bibinfo{author}{Y.~Shen}, \bibinfo{author}{Y.~Zhu},
  \bibinfo{author}{L.~Wang}, \bibinfo{author}{Y.~Wei}, \bibinfo{author}{Y.~Wu},
  \bibinfo{author}{H.~Zeng}, \bibinfo{title}{All Inorganic Halide Perovskites
  Nanosystem: Synthesis, Structural Features, Optical Properties and
  Optoelectronic Applications}, \bibinfo{journal}{Small} \bibinfo{volume}{13}
  (\bibinfo{year}{2017}) \bibinfo{pages}{1603996}.

\bibitem[{Walsh(2015)}]{Walsh15}
\bibinfo{author}{A.~Walsh}, \bibinfo{title}{Principles of Chemical Bonding and
  Band Gap Engineering in Hybrid Organic-Inorganic Halide Perovskites},
  \bibinfo{journal}{J. Phys. Chem. C} \bibinfo{volume}{119}
  (\bibinfo{year}{2015}) \bibinfo{pages}{5755--5760}.

\bibitem[{Saliba et~al.(2016{\natexlab{a}})Saliba, Matsui, Domanski, Seo,
  Ummadisingu, Zakeeruddin, Correa-Baena, Tress, Abate, Hagfeldt, and
  Gr\"{a}tzel}]{Saliba2}
\bibinfo{author}{M.~Saliba}, \bibinfo{author}{T.~Matsui},
  \bibinfo{author}{K.~Domanski}, \bibinfo{author}{J.-Y. Seo},
  \bibinfo{author}{A.~Ummadisingu}, \bibinfo{author}{S.~M. Zakeeruddin},
  \bibinfo{author}{J.-P. Correa-Baena}, \bibinfo{author}{W.~R. Tress},
  \bibinfo{author}{A.~Abate}, \bibinfo{author}{A.~Hagfeldt},
  \bibinfo{author}{M.~Gr\"{a}tzel}, \bibinfo{title}{Incorporation of rubidium
  cations into perovskite solar cells improves photovoltaic performance},
  \bibinfo{journal}{Science} \bibinfo{volume}{354}
  (\bibinfo{year}{2016}{\natexlab{a}}) \bibinfo{pages}{206}.

\bibitem[{Goldschmidt(1926)}]{Goldschmidt}
\bibinfo{author}{V.~M. Goldschmidt}, \bibinfo{title}{Die Gesetze der
  Krystallochemie}, \bibinfo{journal}{Naturwissenschaften} \bibinfo{volume}{14}
  (\bibinfo{year}{1926}) \bibinfo{pages}{477--485}.

\bibitem[{Shannon(1976)}]{Shannon}
\bibinfo{author}{R.~D. Shannon}, \bibinfo{title}{Revised effective ionic-radii
  and systematic studies of interatomic distances in halides and
  chalcogenides}, \bibinfo{journal}{Acta Crystallogr. A} \bibinfo{volume}{32}
  (\bibinfo{year}{1976}) \bibinfo{pages}{751--767}.

\bibitem[{Hendon et~al.(2015)Hendon, Yang, Burton, and Walsh}]{Hendon}
\bibinfo{author}{C.~H. Hendon}, \bibinfo{author}{R.~X. Yang},
  \bibinfo{author}{L.~A. Burton}, \bibinfo{author}{A.~Walsh},
  \bibinfo{title}{Assessment of polyanion (\ce{BF4-} and \ce{PF6-})
  substitutions in hybrid halide perovskites}, \bibinfo{journal}{J. Mater.
  Chem. A} \bibinfo{volume}{3} (\bibinfo{year}{2015})
  \bibinfo{pages}{9067--9070}.

\bibitem[{Hoffman et~al.(2016)Hoffman, Schleper, and Kamat}]{Hoffman}
\bibinfo{author}{J.~B. Hoffman}, \bibinfo{author}{A.~L. Schleper},
  \bibinfo{author}{P.~V. Kamat}, \bibinfo{title}{Transformation of Sintered
  \ce{CsPbBr3} Nanocrystals to Cubic \ce{CsPbI3} and Gradient
  \ce{CsPbBr_xI_{3−x}} through Halide Exchange}, \bibinfo{journal}{J. Am.
  Chem. Soc.} \bibinfo{volume}{138} (\bibinfo{year}{2016})
  \bibinfo{pages}{8603--8611}.

\bibitem[{Kulbak et~al.(2015)Kulbak, Cahen, and Hodes}]{Kulbak}
\bibinfo{author}{M.~Kulbak}, \bibinfo{author}{D.~Cahen},
  \bibinfo{author}{G.~Hodes}, \bibinfo{journal}{J. Phys. Chem. Lett.}
  \bibinfo{volume}{6} (\bibinfo{year}{2015}) \bibinfo{pages}{2452--2456}.

\bibitem[{Kovalsky et~al.(2017)Kovalsky, Wang, Marek, Burda, and
  Dyck}]{Kovalsky}
\bibinfo{author}{A.~Kovalsky}, \bibinfo{author}{L.~Wang},
  \bibinfo{author}{G.~T. Marek}, \bibinfo{author}{C.~Burda},
  \bibinfo{author}{J.~S. Dyck}, \bibinfo{title}{Thermal Conductivity of
  \ce{CH3NH3PbI3} and \ce{CsPbI3}: Measuring the Effect of the Methylammonium
  Ion on Phonon Scattering}, \bibinfo{journal}{J. Phys. Chem. C}
  \bibinfo{volume}{121} (\bibinfo{year}{2017}) \bibinfo{pages}{3228--3233}.

\bibitem[{Eperon et~al.(2015)Eperon, Patern\'{o}, Sutton, Zampetti,
  Haghighirad, Cacialli, and Snaith}]{Eperon}
\bibinfo{author}{G.~E. Eperon}, \bibinfo{author}{G.~M. Patern\'{o}},
  \bibinfo{author}{R.~J. Sutton}, \bibinfo{author}{A.~Zampetti},
  \bibinfo{author}{A.~A. Haghighirad}, \bibinfo{author}{F.~Cacialli},
  \bibinfo{author}{H.~J. Snaith}, \bibinfo{title}{Inorganic caesium lead iodide
  perovskite solar cells}, \bibinfo{journal}{J. Mater. Chem. A}
  \bibinfo{volume}{3} (\bibinfo{year}{2015}) \bibinfo{pages}{19688--19695}.

\bibitem[{Ahmad et~al.(2017)Ahmad, Khan, Niu, and Tang}]{Ahmad}
\bibinfo{author}{W.~Ahmad}, \bibinfo{author}{J.~Khan},
  \bibinfo{author}{G.~Niu}, \bibinfo{author}{J.~Tang},
  \bibinfo{title}{Inorganic \ce{CsPbI3} Perovskite-Based Solar Cells: A Choice
  for a Tandem Device}, \bibinfo{journal}{Sol. RRL} \bibinfo{volume}{1}
  (\bibinfo{year}{2017}) \bibinfo{pages}{1700048}.

\bibitem[{Heidrich et~al.(1981)Heidrich, Schafer, Schreiber, Sochtig, Trendel,
  Treusch, Grandke, and Stolz}]{Heidrich}
\bibinfo{author}{K.~Heidrich}, \bibinfo{author}{W.~Schafer},
  \bibinfo{author}{M.~Schreiber}, \bibinfo{author}{J.~Sochtig},
  \bibinfo{author}{G.~Trendel}, \bibinfo{author}{J.~Treusch},
  \bibinfo{author}{T.~Grandke}, \bibinfo{author}{H.~J. Stolz},
  \bibinfo{journal}{Phys. Rev. B} \bibinfo{volume}{24} (\bibinfo{year}{1981})
  \bibinfo{pages}{5642}.

\bibitem[{Moller(1958)}]{Moller}
\bibinfo{author}{C.~K. Moller}, \bibinfo{title}{Crystal Structure and
  Photoconductivity of Caesium Plumbohalides}, \bibinfo{journal}{Nature}
  \bibinfo{volume}{182} (\bibinfo{year}{1958}) \bibinfo{pages}{1436}.

\bibitem[{Bekenstein et~al.(2015)Bekenstein, Koscher, Eaton, Yang, and
  Alivisatos}]{Bekenstein}
\bibinfo{author}{Y.~Bekenstein}, \bibinfo{author}{B.~A. Koscher},
  \bibinfo{author}{S.~W. Eaton}, \bibinfo{author}{P.~Yang},
  \bibinfo{author}{A.~P. Alivisatos}, \bibinfo{title}{Highly Luminescent
  Colloidal Nanoplates of Perovskite Cesium Lead Halide and Their Oriented
  Assemblies}, \bibinfo{journal}{J. Am. Chem. Soc.} \bibinfo{volume}{137}
  (\bibinfo{year}{2015}) \bibinfo{pages}{16008--16011}.

\bibitem[{Swarnkar et~al.(2016)Swarnkar, Marshall, Sanehira, Chernomordik,
  Moore, Christians, Chakrabarti, and Luther}]{Swarnkar}
\bibinfo{author}{A.~Swarnkar}, \bibinfo{author}{A.~R. Marshall},
  \bibinfo{author}{E.~M. Sanehira}, \bibinfo{author}{B.~D. Chernomordik},
  \bibinfo{author}{D.~T. Moore}, \bibinfo{author}{J.~A. Christians},
  \bibinfo{author}{T.~Chakrabarti}, \bibinfo{author}{J.~M. Luther},
  \bibinfo{title}{Quantum dot−induced phase stabilization of
  $\alpha$-\ce{CsPbI3} perovskite for high-efficiency photovoltaics},
  \bibinfo{journal}{Science} \bibinfo{volume}{354} (\bibinfo{year}{2016})
  \bibinfo{pages}{92--95}.

\bibitem[{Davis et~al.(2017)Davis, de~la Pe\'{n}a, Tabachnyk, Richter, Lamboll,
  Booker, Rivarola, Griffiths, Ducati, Menke, Deschler, and Greenham}]{Davis}
\bibinfo{author}{N.~J. L.~K. Davis}, \bibinfo{author}{F.~J. de~la Pe\'{n}a},
  \bibinfo{author}{M.~Tabachnyk}, \bibinfo{author}{J.~M. Richter},
  \bibinfo{author}{R.~D. Lamboll}, \bibinfo{author}{E.~P. Booker},
  \bibinfo{author}{F.~W.~R. Rivarola}, \bibinfo{author}{J.~T. Griffiths},
  \bibinfo{author}{C.~Ducati}, \bibinfo{author}{S.~M. Menke},
  \bibinfo{author}{F.~Deschler}, \bibinfo{author}{N.~C. Greenham},
  \bibinfo{title}{Photon Reabsorption in Mixed \ce{CsPbCl3}:\ce{CsPbI3}
  Perovskite Nanocrystal Films for Light-Emitting Diodes}, \bibinfo{journal}{J.
  Phys. Chem. C} \bibinfo{volume}{121} (\bibinfo{year}{2017})
  \bibinfo{pages}{3790--3796}.

\bibitem[{Akkerman et~al.(2015)Akkerman, D'Innocenzo, Accornero, Scarpellini,
  Petrozza, Prato, and Manna}]{Akkerman}
\bibinfo{author}{Q.~A. Akkerman}, \bibinfo{author}{V.~D'Innocenzo},
  \bibinfo{author}{S.~Accornero}, \bibinfo{author}{A.~Scarpellini},
  \bibinfo{author}{A.~Petrozza}, \bibinfo{author}{M.~Prato},
  \bibinfo{author}{L.~Manna}, \bibinfo{title}{Tuning the Optical Properties of
  Cesium Lead Halide Perovskite Nanocrystals by Anion Exchange Reactions},
  \bibinfo{journal}{J. Am. Chem. Soc.} \bibinfo{volume}{137}
  (\bibinfo{year}{2015}) \bibinfo{pages}{10276--10281}.

\bibitem[{Nedelcu et~al.(2015)Nedelcu, Protesescu, Yakunin, Bodnarchuk,
  Grotevent, and Kovalenko}]{Nedelcu}
\bibinfo{author}{G.~Nedelcu}, \bibinfo{author}{L.~Protesescu},
  \bibinfo{author}{S.~Yakunin}, \bibinfo{author}{M.~I. Bodnarchuk},
  \bibinfo{author}{M.~J. Grotevent}, \bibinfo{author}{M.~V. Kovalenko},
  \bibinfo{title}{Fast Anion-Exchange in Highly Luminescent Nanocrystals of
  Cesium Lead Halide Perovskites (\ce{CsPbX3}, X = Cl, Br, I)},
  \bibinfo{journal}{Nano Lett.} \bibinfo{volume}{15} (\bibinfo{year}{2015})
  \bibinfo{pages}{5635--5640}.

\bibitem[{Gomez et~al.(2017)Gomez, de~Weerd, Huesob, and
  Gregorkiewicza}]{Gomez}
\bibinfo{author}{L.~Gomez}, \bibinfo{author}{C.~de~Weerd},
  \bibinfo{author}{J.~L. Huesob}, \bibinfo{author}{T.~Gregorkiewicza},
  \bibinfo{title}{Color-stable water-dispersed cesium lead halide perovskite
  nanocrystals}, \bibinfo{journal}{Nanoscale} \bibinfo{volume}{9}
  (\bibinfo{year}{2017}) \bibinfo{pages}{631--636}.

\bibitem[{Liu et~al.(2016)Liu, Wang, Sui, Wang, Chi, Wang, Chen, Ji, Zou, and
  Zhang}]{QLiu}
\bibinfo{author}{Q.~Liu}, \bibinfo{author}{Y.~Wang}, \bibinfo{author}{N.~Sui},
  \bibinfo{author}{Y.~Wang}, \bibinfo{author}{X.~Chi},
  \bibinfo{author}{Q.~Wang}, \bibinfo{author}{Y.~Chen},
  \bibinfo{author}{W.~Ji}, \bibinfo{author}{L.~Zou},
  \bibinfo{author}{H.~Zhang}, \bibinfo{title}{Exciton Relaxation Dynamics in
  Photo-Excited \ce{CsPbI3} Perovskite Nanocrystals}, \bibinfo{journal}{Sci.
  Rep.} \bibinfo{volume}{6} (\bibinfo{year}{2016}) \bibinfo{pages}{29442}.

\bibitem[{Protesescu et~al.(2015)Protesescu, Yakunin, Bodnarchuk, Krieg,
  Caputo, Hendon, Yang, Walsh, and Kovalenko}]{Protesescu}
\bibinfo{author}{L.~Protesescu}, \bibinfo{author}{S.~Yakunin},
  \bibinfo{author}{M.~I. Bodnarchuk}, \bibinfo{author}{F.~Krieg},
  \bibinfo{author}{R.~Caputo}, \bibinfo{author}{C.~H. Hendon},
  \bibinfo{author}{R.~X. Yang}, \bibinfo{author}{A.~Walsh},
  \bibinfo{author}{M.~V. Kovalenko}, \bibinfo{title}{Nanocrystals of Cesium
  Lead Halide Perovskites (\ce{CsPbX3}, X = Cl, Br, and I): Novel
  Optoelectronic Materials Showing Bright Emission with Wide Color Gamut},
  \bibinfo{journal}{Nano Lett.} \bibinfo{volume}{15} (\bibinfo{year}{2015})
  \bibinfo{pages}{3692--3696}.

\bibitem[{Li et~al.(2016{\natexlab{b}})Li, Yang, Park, Wei, Berry, and
  Zhu}]{ZLi}
\bibinfo{author}{Z.~Li}, \bibinfo{author}{M.~Yang}, \bibinfo{author}{J.-S.
  Park}, \bibinfo{author}{S.-H. Wei}, \bibinfo{author}{J.~J. Berry},
  \bibinfo{author}{K.~Zhu}, \bibinfo{title}{Stabilizing perovskite structures
  by tuning tolerance factor: formation of formamidinium and cesium lead iodide
  solid-state alloys}, \bibinfo{journal}{Chem. Mater.} \bibinfo{volume}{28}
  (\bibinfo{year}{2016}{\natexlab{b}}) \bibinfo{pages}{284--292}.

\bibitem[{Niu et~al.(2017)Niu, Li, Li, Liang, and Wang}]{Niu}
\bibinfo{author}{G.~Niu}, \bibinfo{author}{W.~Li}, \bibinfo{author}{J.~Li},
  \bibinfo{author}{X.~Liang}, \bibinfo{author}{L.~Wang},
  \bibinfo{title}{Enhancement of thermal stability for perovskite solar cells
  through cesium doping}, \bibinfo{journal}{RSC Adv.} \bibinfo{volume}{7}
  (\bibinfo{year}{2017}) \bibinfo{pages}{17473--17479}.

\bibitem[{Lee et~al.(2015)Lee, Kim, Kim, Seo, Cho, and Park}]{Lee}
\bibinfo{author}{J.-W. Lee}, \bibinfo{author}{D.-H. Kim},
  \bibinfo{author}{H.-S. Kim}, \bibinfo{author}{S.-W. Seo},
  \bibinfo{author}{S.~M. Cho}, \bibinfo{author}{N.-G. Park},
  \bibinfo{title}{Formamidinium and Cesium Hybridization for Photo- and
  Moisture-Stable Perovskite Solar Cell}, \bibinfo{journal}{Adv. Energy Mater.}
  \bibinfo{volume}{5} (\bibinfo{year}{2015}) \bibinfo{pages}{1501310}.

\bibitem[{McMeekin et~al.(2016)McMeekin, Sadoughi, Rehman, Eperon, Saliba,
  Horantner, Haghighirad, Sakai, Korte, Rech, Johnston, Herz, and
  Snaith}]{McMeekin}
\bibinfo{author}{D.~P. McMeekin}, \bibinfo{author}{G.~Sadoughi},
  \bibinfo{author}{W.~Rehman}, \bibinfo{author}{G.~E. Eperon},
  \bibinfo{author}{M.~Saliba}, \bibinfo{author}{M.~T. Horantner},
  \bibinfo{author}{A.~Haghighirad}, \bibinfo{author}{N.~Sakai},
  \bibinfo{author}{L.~Korte}, \bibinfo{author}{B.~Rech}, \bibinfo{author}{M.~B.
  Johnston}, \bibinfo{author}{L.~M. Herz}, \bibinfo{author}{H.~J. Snaith},
  \bibinfo{title}{A mixed-cation lead mixed-halide perovskite absorber for
  tandem solar cells}, \bibinfo{journal}{Science} \bibinfo{volume}{351}
  (\bibinfo{year}{2016}) \bibinfo{pages}{151--154}.

\bibitem[{Rehman et~al.(2017)Rehman, McMeekin, Patel, Milot, Johnston, Snaith,
  and Herz}]{Rehman}
\bibinfo{author}{W.~Rehman}, \bibinfo{author}{D.~P. McMeekin},
  \bibinfo{author}{J.~B. Patel}, \bibinfo{author}{R.~L. Milot},
  \bibinfo{author}{M.~B. Johnston}, \bibinfo{author}{H.~J. Snaith},
  \bibinfo{author}{L.~M. Herz}, \bibinfo{title}{Photovoltaic mixed-cation lead
  mixed-halide perovskites: links between crystallinity, photo-stability and
  electronic properties}, \bibinfo{journal}{Energy Environ. Sci.}
  \bibinfo{volume}{10} (\bibinfo{year}{2017}) \bibinfo{pages}{361--369}.

\bibitem[{Saliba et~al.(2016{\natexlab{b}})Saliba, Matsui, Seo, Domanski,
  Correa-Baena, Nazeeruddin, Zakeeruddin, Tress, Abate, Hagfeldtd, and
  Gr\"{a}tzel}]{Saliba}
\bibinfo{author}{M.~Saliba}, \bibinfo{author}{T.~Matsui},
  \bibinfo{author}{J.-Y. Seo}, \bibinfo{author}{K.~Domanski},
  \bibinfo{author}{J.-P. Correa-Baena}, \bibinfo{author}{M.~K. Nazeeruddin},
  \bibinfo{author}{S.~M. Zakeeruddin}, \bibinfo{author}{W.~Tress},
  \bibinfo{author}{A.~Abate}, \bibinfo{author}{A.~Hagfeldtd},
  \bibinfo{author}{M.~Gr\"{a}tzel}, \bibinfo{title}{Cesium-containing triple
  cation perovskite solar cells: improved stability, reproducibility and high
  efficiency}, \bibinfo{journal}{Energy Environ. Sci.} \bibinfo{volume}{9}
  (\bibinfo{year}{2016}{\natexlab{b}}) \bibinfo{pages}{1989--1997}.

\bibitem[{Zhang et~al.(2017)Zhang, Yun, Ma, Zheng, Lau, Deng, Kim, Kim, Seidel,
  Green, Huang, and Ho-Baillie}]{Zhang}
\bibinfo{author}{M.~Zhang}, \bibinfo{author}{J.~S. Yun},
  \bibinfo{author}{Q.~Ma}, \bibinfo{author}{J.~Zheng},
  \bibinfo{author}{C.~F.~J. Lau}, \bibinfo{author}{X.~Deng},
  \bibinfo{author}{J.~Kim}, \bibinfo{author}{D.~Kim},
  \bibinfo{author}{J.~Seidel}, \bibinfo{author}{M.~A. Green},
  \bibinfo{author}{S.~Huang}, \bibinfo{author}{A.~W.~Y. Ho-Baillie},
  \bibinfo{title}{High-Efficiency Rubidium-Incorporated Perovskite Solar Cells
  by Gas Quenching}, \bibinfo{journal}{ACS Energy Lett.} \bibinfo{volume}{2}
  (\bibinfo{year}{2017}) \bibinfo{pages}{438--444}.

\bibitem[{Duong et~al.(2017)Duong, Wu, Shen, Peng, Fu, Jacobs, Wang, Kho, Fong,
  Stocks, Franklin, Blakers, and \textit{et al.}}]{Duong}
\bibinfo{author}{T.~Duong}, \bibinfo{author}{Y.~L. Wu},
  \bibinfo{author}{H.~Shen}, \bibinfo{author}{J.~Peng},
  \bibinfo{author}{X.~Fu}, \bibinfo{author}{D.~Jacobs}, \bibinfo{author}{E.-C.
  Wang}, \bibinfo{author}{T.~C. Kho}, \bibinfo{author}{K.~C. Fong},
  \bibinfo{author}{M.~Stocks}, \bibinfo{author}{E.~Franklin},
  \bibinfo{author}{A.~Blakers}, \bibinfo{author}{\textit{et al.}},
  \bibinfo{title}{Rubidium Multication Perovskite with Optimized Bandgap for
  Perovskite-Silicon Tandem with over 26\% Efficiency}, \bibinfo{journal}{Adv.
  Energy Mater.}  (\bibinfo{year}{2017}) \bibinfo{pages}{1700228}.

\bibitem[{Brgoch et~al.(2014)Brgoch, Lehner, Chabinyc, and Seshadri}]{Brgoch}
\bibinfo{author}{J.~Brgoch}, \bibinfo{author}{A.~J. Lehner},
  \bibinfo{author}{M.~Chabinyc}, \bibinfo{author}{R.~Seshadri},
  \bibinfo{title}{Ab Initio Calculations of Band Gaps and Absolute Band
  Positions of Polymorphs of \ce{RbPbI3} and \ce{CsPbI3}: Implications for
  Main-Group Halide Perovskite Photovoltaics}, \bibinfo{journal}{J. Phys. Chem.
  C} \bibinfo{volume}{118} (\bibinfo{year}{2014})
  \bibinfo{pages}{27721--27727}.

\bibitem[{Murtaza and Ahmad(2011)}]{Murtaza}
\bibinfo{author}{G.~Murtaza}, \bibinfo{author}{I.~Ahmad}, \bibinfo{title}{First
  principle study of the structural and optoelectronic properties of cubic
  perovskites \ce{CsPbM3} (M = Cl,Br,I)}, \bibinfo{journal}{Physica B}
  \bibinfo{volume}{206} (\bibinfo{year}{2011}) \bibinfo{pages}{3222--3229}.

\bibitem[{Afsari et~al.(2016)Afsari, Boochani, and Hantezadeh}]{Afsari}
\bibinfo{author}{M.~Afsari}, \bibinfo{author}{A.~Boochani},
  \bibinfo{author}{M.~Hantezadeh}, \bibinfo{title}{Electronic, optical and
  elastic properties of cubic perovskite CsPbI3: Using first principles study},
  \bibinfo{journal}{Optik} \bibinfo{volume}{127} (\bibinfo{year}{2016})
  \bibinfo{pages}{11433--11443}.

\bibitem[{Yang et~al.(2017{\natexlab{b}})Yang, Skelton, da~Silva, Frost, and
  Walsh}]{RYang}
\bibinfo{author}{R.~X. Yang}, \bibinfo{author}{J.~M. Skelton},
  \bibinfo{author}{E.~L. da~Silva}, \bibinfo{author}{J.~M. Frost},
  \bibinfo{author}{A.~Walsh}, \bibinfo{title}{Spontaneous Octahedral Tilting in
  the Cubic Inorganic Caesium Halide Perovskites \ce{CsSnX3} and \ce{CsPbX3} (X
  = F, Cl, Br, I)}  (\bibinfo{year}{2017}{\natexlab{b}})
  \bibinfo{pages}{arXiv:1708.00499}.

\bibitem[{Song et~al.(2017)Song, Gao, Li, and Zhang}]{GSong}
\bibinfo{author}{G.~Song}, \bibinfo{author}{B.~Gao}, \bibinfo{author}{G.~Li},
  \bibinfo{author}{J.~Zhang}, \bibinfo{title}{First-principles study on the
  electric structure and ferroelectricity in epitaxial \ce{CsSnI3} films},
  \bibinfo{journal}{RSC Adv.} \bibinfo{volume}{7} (\bibinfo{year}{2017})
  \bibinfo{pages}{41077--41083}.

\bibitem[{da~Silva et~al.(2015)da~Silva, Skelton, Parker, and Walsh}]{Silva}
\bibinfo{author}{E.~L. da~Silva}, \bibinfo{author}{J.~M. Skelton},
  \bibinfo{author}{S.~C. Parker}, \bibinfo{author}{A.~Walsh},
  \bibinfo{title}{Phase stability and transformations in the halide perovskite
  \ce{CsSnI3}}, \bibinfo{journal}{Phys. Rev. B} \bibinfo{volume}{91}
  (\bibinfo{year}{2015}) \bibinfo{pages}{144107}.

\bibitem[{yi~Huang and Lambrecht(2013)}]{Huang}
\bibinfo{author}{L.~yi~Huang}, \bibinfo{author}{W.~R.~L. Lambrecht},
  \bibinfo{title}{Electronic band structure, phonons, and exciton binding
  energies of halide perovskites \ce{CsSnCl3}, \ce{CsSnBr3}, and \ce{CsSnI3}},
  \bibinfo{journal}{Phys. Rev. B} \bibinfo{volume}{88} (\bibinfo{year}{2013})
  \bibinfo{pages}{165203}.

\bibitem[{{P. Giannozzi and S. Baroni and N. Bonini and M. Calandra and R. Car,
  {\it et al.}}(2009)}]{QE}
\bibinfo{author}{{P. Giannozzi and S. Baroni and N. Bonini and M. Calandra and
  R. Car, {\it et al.}}}, \bibinfo{title}{{QUANTUM ESPRESSO}: a modular and
  open-source software project for quantum simulations of materials},
  \bibinfo{journal}{J. Phys.:Condens. Matter} \bibinfo{volume}{21}
  (\bibinfo{year}{2009}) \bibinfo{pages}{395502}.

\bibitem[{Yu and Emmerich(2007)}]{yucj07}
\bibinfo{author}{C.-J. Yu}, \bibinfo{author}{H.~Emmerich}, \bibinfo{title}{An
  efficient virtual crystal approximation that can be used to treat
  heterovalent atoms, applied to $(1-x)$\ce{BiScO3}-$x$\ce{PbTiO3}},
  \bibinfo{journal}{J. Phys.: Condens. Matter} \bibinfo{volume}{19}
  (\bibinfo{year}{2007}) \bibinfo{pages}{306203}.

\bibitem[{Jong et~al.(2016)Jong, Yu, Ri, Kim, and Ri}]{yucj10}
\bibinfo{author}{U.-G. Jong}, \bibinfo{author}{C.-J. Yu},
  \bibinfo{author}{J.-S. Ri}, \bibinfo{author}{N.-H. Kim},
  \bibinfo{author}{G.-C. Ri}, \bibinfo{title}{Influence of halide composition
  on the structural, electronic, and optical properties of mixed
  \ce{CH3NH3Pb(I_{1-x}Br_x)3} perovskites calculated using the virtual crystal
  approximation method}, \bibinfo{journal}{Phys. Rev. B} \bibinfo{volume}{94}
  (\bibinfo{year}{2016}) \bibinfo{pages}{125139}.

\bibitem[{Jong et~al.(2017)Jong, Yu, Jang, Ri, Hong, and Pae}]{yucj12}
\bibinfo{author}{U.-G. Jong}, \bibinfo{author}{C.-J. Yu},
  \bibinfo{author}{Y.-M. Jang}, \bibinfo{author}{G.-C. Ri},
  \bibinfo{author}{S.-N. Hong}, \bibinfo{author}{Y.-H. Pae},
  \bibinfo{title}{Revealing the stability and efficiency enhancement in mixed
  halide perovskites \ce{MAPb(I_{1-x}Cl_x)3} with ab initio calculations},
  \bibinfo{journal}{J. Power Sources} \bibinfo{volume}{350}
  (\bibinfo{year}{2017}) \bibinfo{pages}{65--72}.

\bibitem[{Birch(1947)}]{birch}
\bibinfo{author}{F.~Birch}, \bibinfo{title}{Finite elastic strain of cubic
  crystal}, \bibinfo{journal}{Phys. Rev.} \bibinfo{volume}{71}
  (\bibinfo{year}{1947}) \bibinfo{pages}{809--824}.

\bibitem[{Trots and Myagkota(2008)}]{Trots}
\bibinfo{author}{D.~M. Trots}, \bibinfo{author}{S.~V. Myagkota},
  \bibinfo{title}{High-Temperature Structural Evolution of Caesium and Rubidium
  Triiodoplumbates}, \bibinfo{journal}{J. Phys. Chem. Solids}
  \bibinfo{volume}{69} (\bibinfo{year}{2008}) \bibinfo{pages}{2520--2526}.

\bibitem[{Sharma et~al.(2003)Sharma, Dewhurst, and Ambrosch-Draxl}]{Sharma}
\bibinfo{author}{S.~Sharma}, \bibinfo{author}{J.~K. Dewhurst},
  \bibinfo{author}{C.~Ambrosch-Draxl}, \bibinfo{journal}{Phys. Rev. B}
  \bibinfo{volume}{67} (\bibinfo{year}{2003}) \bibinfo{pages}{165332}.

\bibitem[{Yu et~al.(2016)Yu, Jong, Ri, Ri, and Pae}]{yucj06}
\bibinfo{author}{C.-J. Yu}, \bibinfo{author}{U.-G. Jong},
  \bibinfo{author}{M.-H. Ri}, \bibinfo{author}{G.-C. Ri},
  \bibinfo{author}{Y.-H. Pae}, \bibinfo{title}{Electronic structure and
  photoabsorption property of pseudocubic perovskites \ce{CH3NH3PbX3} (X = I,
  Br) including van der Waals interaction}, \bibinfo{journal}{J. Mater. Sci.}
  \bibinfo{volume}{51} (\bibinfo{year}{2016}) \bibinfo{pages}{9849--9854}.

\bibitem[{Kresse and Furthm\"{u}ller(1996)}]{vasp1}
\bibinfo{author}{G.~Kresse}, \bibinfo{author}{J.~Furthm\"{u}ller},
  \bibinfo{title}{Effciency of Ab-initio Total Energy Calculations for Metals
  and Semiconductors Using a Plane-wave Basis Set}, \bibinfo{journal}{Comput.
  Mater. Sci.} \bibinfo{volume}{6} (\bibinfo{year}{1996})
  \bibinfo{pages}{15--50}.

\bibitem[{Perdew et~al.(1996)Perdew, Burke, and Ernzerhof}]{pbe}
\bibinfo{author}{J.~P. Perdew}, \bibinfo{author}{K.~Burke},
  \bibinfo{author}{M.~Ernzerhof}, \bibinfo{title}{Generalized Gradient
  Approximation Made Simple}, \bibinfo{journal}{Phys. Rev. Lett.}
  \bibinfo{volume}{77} (\bibinfo{year}{1996}) \bibinfo{pages}{3865}.

\bibitem[{Even et~al.(2013)Even, Pedesseau, Jancu, and Katan}]{Even13}
\bibinfo{author}{J.~Even}, \bibinfo{author}{L.~Pedesseau},
  \bibinfo{author}{J.-M. Jancu}, \bibinfo{author}{C.~Katan},
  \bibinfo{title}{Importance of spin-orbit coupling in hybrid organic/inorganic
  perovskites for photovoltaic applications}, \bibinfo{journal}{J. Phys. Chem.
  Lett.} \bibinfo{volume}{4} (\bibinfo{year}{2013})
  \bibinfo{pages}{2999--3005}.

\bibitem[{Perdew et~al.(2008)Perdew, Ruzsinszky, Csonka, Vydrov, Scuseria,
  Constantin, Zhou, and Burke}]{PBEsol}
\bibinfo{author}{J.~P. Perdew}, \bibinfo{author}{A.~Ruzsinszky},
  \bibinfo{author}{G.~I. Csonka}, \bibinfo{author}{O.~A. Vydrov},
  \bibinfo{author}{G.~E. Scuseria}, \bibinfo{author}{L.~A. Constantin},
  \bibinfo{author}{X.~Zhou}, \bibinfo{author}{K.~Burke},
  \bibinfo{title}{Restoring the Density Gradient Expansion for Exchange in
  Solids and Surfaces}, \bibinfo{journal}{Phys. Rev. Lett.}
  \bibinfo{volume}{100} (\bibinfo{year}{2008}) \bibinfo{pages}{136406}.

\bibitem[{Lee et~al.(2010)Lee, Murray, Kong, Lundqvist, and Langreth}]{vdwDF2}
\bibinfo{author}{K.~Lee}, \bibinfo{author}{E.~D. Murray},
  \bibinfo{author}{L.~Kong}, \bibinfo{author}{B.~I. Lundqvist},
  \bibinfo{author}{D.~C. Langreth}, \bibinfo{title}{High-accuracy van der Waals
  density functional}, \bibinfo{journal}{Phys. Rev. B} \bibinfo{volume}{82}
  (\bibinfo{year}{2010}) \bibinfo{pages}{081101(R)}.

\bibitem[{Umari et~al.(2014)Umari, Mosconi, and Angelis}]{Umari}
\bibinfo{author}{P.~Umari}, \bibinfo{author}{E.~Mosconi},
  \bibinfo{author}{F.~D. Angelis}, \bibinfo{title}{Relativistic GW calculations
  on \ce{CH3NH3PbI3} and \ce{CH3NH3SnI3} Perovskites for Solar Cell
  Applications}, \bibinfo{journal}{Sci. Rep.} \bibinfo{volume}{4}
  (\bibinfo{year}{2014}) \bibinfo{pages}{4467}.

\bibitem[{Atourki et~al.(2016)Atourki, Vega, Mar\`{i}, Mollarb, Ahsainec,
  Bouabida, and Ihlal}]{Atourki}
\bibinfo{author}{L.~Atourki}, \bibinfo{author}{E.~Vega},
  \bibinfo{author}{B.~Mar\`{i}}, \bibinfo{author}{M.~Mollarb},
  \bibinfo{author}{H.~A. Ahsainec}, \bibinfo{author}{K.~Bouabida},
  \bibinfo{author}{A.~Ihlal}, \bibinfo{title}{Role of the chemical substitution
  on the structural and luminescence properties of the mixed halide perovskite
  thin MAPbI$_{3-x}$Br$_x$ ($0<x<1$) films}, \bibinfo{journal}{Appl. Surf.
  Sci.} \bibinfo{volume}{371} (\bibinfo{year}{2016}) \bibinfo{pages}{112--117}.

\end{thebibliography}

\end{document}